\documentclass[twocolumn,twoside,slac_two]{revtex4}
\pdfoutput=1
\usepackage{graphicx}
\usepackage{fancyhdr}
\usepackage{unites2e}
\begin{document}

\title{Galactic Binary Systems}

%

\author{J. Holder}
\affiliation{Department of Physics and the Bartol Research Institute, University of Delaware, DE 19716, USA}
\begin{abstract}
The population of binary systems known to emit in the GeV and TeV
bands consists of only a few firmly identified Galactic sources. These
rare objects constitute extreme particle accelerators operating under
varying, but regularly repeating, conditions. As such, they provide
access to a unique laboratory in which to study particle acceleration,
and the nature of gamma-ray production, emission and absorption
processes near compact objects. Here we review the current
observational status of the field, and discuss some of the recent
interpretations of the results.

\end{abstract}

\maketitle

\thispagestyle{fancy}


\section{Introduction}

After many decades as an experimental technique at the fringes of
mainstream astronomy, gamma-ray observations now constitute a
well-established astronomical discipline. The Fermi-LAT catalogue is
expected to increase the number of known high-energy (HE;
$30\U{MeV}-30\U{GeV}$) sources into the many thousands, while the
number of very-high-energy (VHE; $30\U{GeV}-30\U{TeV}$) sources is
approaching 100. This allows for unbiased population studies of the
major source classes (blazars, pulsars, pulsar wind nebulae, supernova
remnants, etc.), which can probe the details of their evolution and
the impact of their surrounding environments. Within this
proliferation of objects, Galactic binary systems known to emit in the
HE/VHE bands constitute just 6 sources, including one marginal
detection and another which may not be a binary system at all. Their
impact on high energy astrophysics is, however, disproportionately
large.

The known gamma-ray binary systems comprise a compact object (black
hole or neutron star) orbiting a massive (O or B type) main sequence
companion. As such, they constitute an astrophysical particle
accelerator operating under a varying, but regularly repeating, set of
environmental conditions. Throughout the orbit, matter and photon
field densities, as well as the system geometry and orientation with
respect to the observer's line of sight are continually changing. In
addition, each system is unique; providing a range of stellar
properties, compact object masses and orbital ephemerides. As a
result, observations of gamma-ray binary systems can provide stringent
and repeatable tests for models of particle acceleration and high
energy emission in extreme astrophysical environments. Furthermore,
the study of Galactic binary systems containing accretion powered jets
may shed light upon the general mechanisms for astrophysical jet
formation, with application to the much larger-scale structures
produced in active galaxies.

In this report we review the known Galactic gamma-ray binary systems,
with a focus on the current observational status in the GeV - TeV
energy region.

\section{History}

\begin{figure*}
\centering
\includegraphics[width=150mm]{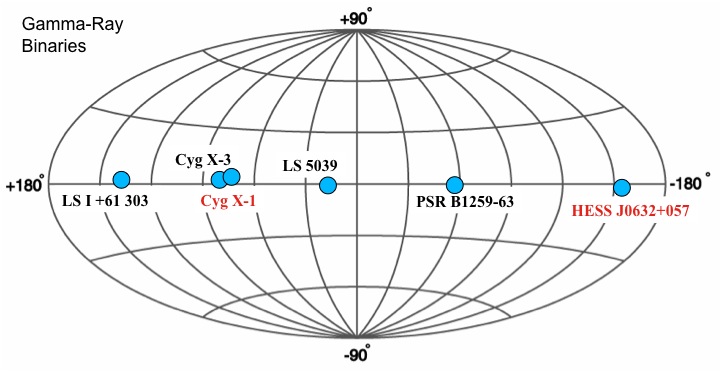}
\caption{The location of gamma-ray sources which might be associated
with binary systems. LS~5039 and LS~I~+61${^\circ}$303 are well
established HE and VHE sources, PSR~B1259-63/SS~2883 is a clear VHE
source. Cyg~X-1 has been detected at $4\sigma$ during one flaring
episode with MAGIC (VHE), Cyg~X-3 is only detected at
HE. HESS~J0632+057 is only detected at VHE, and is not firmly
identified as a binary system.} \label{map}
\end{figure*}

To some extent, the whole field of VHE gamma-ray astronomy owes its
existence to the study of binary systems. Observations close to a huge
radio flare of Cygnus X-3 in 1972 led to claims of VHE gamma-ray
emission \cite{vladimirsky73}, which were repeated by numerous
Cherenkov telescopes and cosmic ray particle detectors observing over
a wide range of the gamma-ray spectrum during the 1970's and
1980's. Periodic emission from several other X-ray binary systems was
also claimed (see e.g. \cite{chadwick90} for a review), which provided
some of the impetus for the construction of new experiments in the
1990's. These new instruments (notably Whipple, HEGRA and CAT) were
equipped with imaging cameras which allowed to greatly reduce the
background of cosmic ray showers, providing sensitivity at the level
of $\sim5\%$ of the steady Crab Nebula flux. Despite this, they failed
to confirm any of the earlier results.

At the lower energies probed by satellite-borne gamma-ray telescopes,
among the 13 gamma-ray sources detected by COS-B was one, 2CG~135+01,
whose error box contained a periodic radio and X-ray source,
LS~I~+61${^\circ}$303 \cite{swanenburg81}. EGRET, on board the CGRO,
also detected a bright GeV source at this location (3EG~J0241+6103,
\cite{tavani98}), as well as other possible binary associations
3EG~J1824-1514 (LS~5039) \cite{paredes00} and 2EG~J2033+4112 (Cyg~X-3)
\cite{mori97}. In each case, however, there was weak or no evidence
for variability, no clear periodicity, and limited positional
accuracy, which made the associations far from definitive.

The next breakthrough came from the ground, starting in 2004, when the
current generation of atmospheric Cherenkov detectors (H.E.S.S., MAGIC
and VERITAS) began to come online. These instruments provided an order
of magnitude improvement in sensitivity over the previous generation,
along with pointing accuracies of $\sim0.1^{\circ}$. The detections of
VHE emission from PSR~B1259-63/SS~2883 \cite{Aharonian_1259detect},
LS~5039 \cite{Aharonian_LS5039} and LS~I~+61${^\circ}$303
\cite{Albert_LSI_detect} provided the first incontrovertible evidence
for orbitally modulated gamma-ray emission from Galactic binary
systems.

This year's results from Fermi-LAT and AGILE reveal definitive HE
detections of LS~I~+61${^\circ}$303 \cite{abdo09_LSI}, LS~5039
\cite{abdo09_LS5039} and Cyg X-3 \cite{agile09_CygX3, fermi09_CygX3}.
Long-term, sensitive monitoring of Galactic binary sources is
providing key insght into their nature, while the critical importance
of obtaining contemporaneous, time-resolved observations of the
complete non-thermal spectra for these objects is becoming clear.

\section{An overview of high energy binary systems}

Fig.~\ref{map} shows the distribution, in Galactic coordinates, of the
six known gamma-ray sources which have been linked with binary
systems. in the following sections we will provide a brief description
of the systems and review the status of gamma-ray observations.


\subsection{PSR~B1259-63/SS~2883}

\begin{figure}
\centering
\includegraphics[width=75mm]{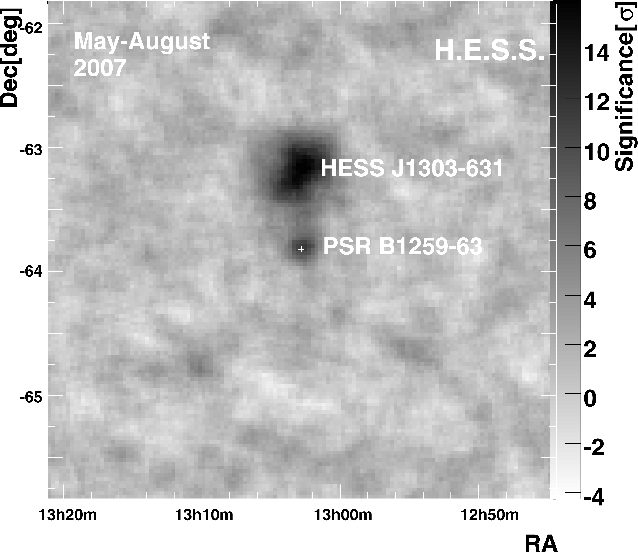}
\caption{Significance map of the VHE gamma-ray excesss for the region
around PSR~B1259-63 (from \cite{Aharonian_1259_2007}).}
\label{PSR1259map}
\end{figure}

PSR~B1259-63/SS~2883 was the first gamma-ray binary system to be
firmly detected at TeV energies, and the first known variable VHE
source in our Galaxy. The system was discovered in the radio
\cite{johnston92}, and comprises a $48\U{ms}$ pulsar orbiting a
massive B2Ve companion. The orbit is highly eccentric ($e=0.87$), with
a period of 3.4 years. At apastron, the orbital separation is
$\sim10\U{A.U}$. At periastron, the separation is only $0.7\U{A.U.}$
and the pulsar passes close to, or possibly through, the circumstellar
disk of the Be star, which is likely inclined with respect to the
plane of the pulsar orbit. A remarkable feature of the initial
H.E.S.S. detection was the discovery of a bright, spatially extended
TeV source (HESS~J1303-631) just $0.6^{\circ}$ to the north of
PSR~B1259-63 (Figure~\ref{PSR1259map}), with no obvious counterpart at
other wavelengths. There are now known to be many similar
unidentified VHE sources along the plane of the Galaxy; the angular
proximity of HESS~J1303-631 and PSR~B1259-63 is purely coincidental.

\begin{figure}
\centering
\includegraphics[width=75mm]{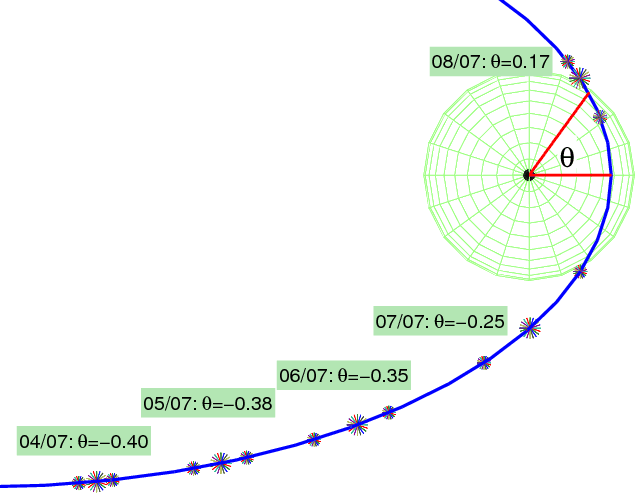}
\caption{A sketch of the PSR~B1259-63/SS2883 orbit (from
\cite{Aharonian_1259_2007}). The circumstellar disk, assumed to extend
out to 20 stellar radii, is shown in green. Dates of
H.E.S.S. observations in 2007 are indicated, together with the value
of the true anomaly, $\theta$, as defined by the red lines.}
\label{PSR1259orbit}
\end{figure}

\begin{figure*}[]
\centering
\includegraphics[width=150mm]{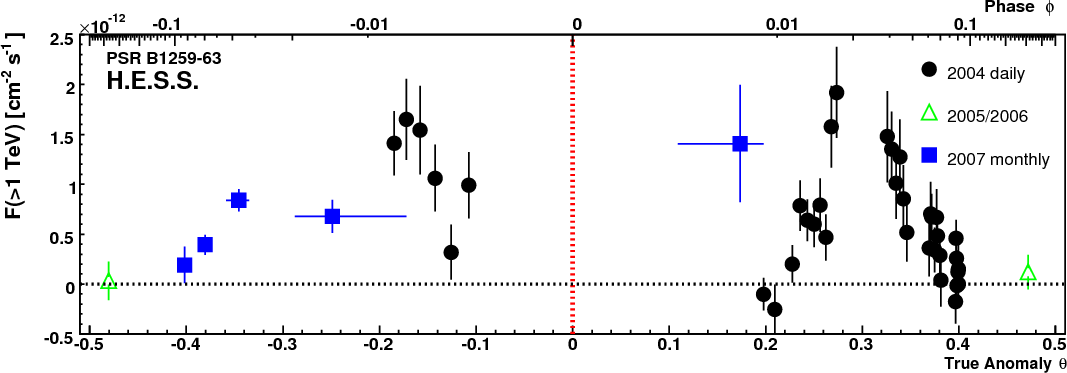}
\caption{Integrated VHE flux above 1\U{TeV} from all combined HESS observations of PSR~B1259-63 as a function of the true anomaly, with the corresponding orbital phases shown on the upper axis (from \cite{Aharonian_1259_2007}).} \label{combined1259}
\end{figure*}

The combination of a long orbit, which is out of phase with our own
orbit around the sun, complicates observations of this
object. Figure~\ref{PSR1259orbit} shows a top view sketch of the
system, including a definition of the true anomaly, $\theta$.
Figure~\ref{combined1259} shows the complete H.E.S.S. dataset plotted
with respect to this true anomaly.  Note that, as yet, due to observing
constraints, there are no observations of the system at
periastron. The TeV emission exhibits two peaks, approximately 15 days
before and after periastron. Within the limited sampling, it also
appears that the lightcurve may be asymmetric with respect to
periastron. The source spectrum at VHE energies shows no sign of
variability, and is well-fit by a power-law with a photon index
$\Gamma=2.8$.

Various authors have attempted to explain the double bumped VHE
lightcurve within a 'hadronic disk scenario', in which the
circumstellar disk provides target material for accelerated hadrons,
leading to $\pi^0$ production and subsequent TeV gamma-ray emission
\cite{kawachi04, chernyakova06}. The newly published 2007
H.E.S.S. observations disfavour this, since the onset of TeV emission
occurs $\sim50\U{days}$ prior to periastron, well before interactions
with the disk could be expected to play a significant role. The next
periastron will occur in December 2010, for the first time with
Fermi-LAT coverage. Accompanying H.E.S.S. observations should also be
possible, particularly during the post-periastron emission period.

\subsection{LS~5039}

\begin{figure}
\centering
\includegraphics[width=75mm]{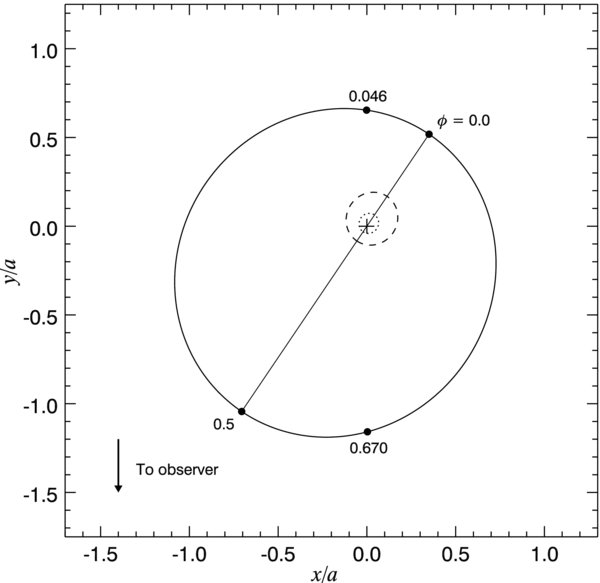}
\caption{A top view of the orbit of LS~5039, showing the relative
orbits (r/a) of the optical star and its compact companion of unknown
mass. The relative orbit of the compact object is shown as a solid
line, while the optical star's relative orbit depends greatly on the mass
of the companion. The dashed line indicates the optical star's orbit
assuming a 3.7 M$_{\odot}$ black hole, while the dotted line assumes a
1.4 M$_{\odot}$ neutron star. Figure from Aragona et
al. \cite{aragona09}.}
\label{LS5039orbit}
\end{figure}

LS~5039 consists of a compact object, either neutron star or black
hole, orbiting a massive O6.5V ($\sim$23~M$_\odot$, \cite{casares05})
star in a 3.9 day orbit. The orbit is slightly eccentric ($e=0.34$),
inclined to the line of sight, and the orbital separation varies from
$\sim0.1\U{A.U.}$ at periastron to $\sim0.2\U{A.U.}$ at
apastron. Figure~\ref{LS5039orbit} sketches the orbit of the system. 

Evidence for high energy emission associated with this source was
first claimed by Paredes et al. \cite{paredes00}, who noted the
coincidence with an EGRET source, 3EG~J1824-1514, and identified
jet-like radio structures, leading to a microquasar
interpretation. Variability in the EGRET source was never established,
however, and high resolution VLBA radio observations by Ribo et
al. \cite{ribo08} show morphological changes on short timescales which
do not support the existence of a persistent jet.

Observations by H.E.S.S. in 2004 revealed that LS~5039 is a bright
source of VHE gamma-rays \cite{Aharonian_LS5039}. Unlike PSR~B1259-63
(and, to a lesser extent, LS~I~+61${^\circ}$303) LS~5039 is almost
perfectly suited to TeV observations, with a short orbital period and
a convenient declination angle, allowing sensitive observations at all
phases over numerous orbits. Deep follow-up observations by
H.E.S.S. provided a measurement of periodic variability in the TeV
emission, modulated at the orbital period. The VHE emission is largely
confined to half of the orbit, peaking around inferior conjunction,
when the compact object is closest to us and co-aligned with our
line-of-sight ($\phi=0.67$ on Figure~\ref{LS5039orbit}). The spectrum
is also orbitally modulated, appearing significantly harder around
inferior conjuction
($\Gamma=1.85\pm0.06_{\mathrm{stat}}\pm0.1_{\mathrm{syst}}$), but with
an exponential cut-off at $E_o=8.7\pm2.0\U{TeV}$. Recent X-ray
observations with Suzaku, which continuously monitored one and a half
full orbits \cite{takahashi09}, show a light curve remarkably similar
to the TeV light curve, but with an amplitude of modulation $\sim3$
times smaller. The hard X-ray spectrum and absence of X-ray emission
lines favours a non-thermal origin for the emission.

\begin{figure}
\vspace{0.5cm}
\centering
\includegraphics[width=75mm, trim=0 0 0 137mm, clip=true]{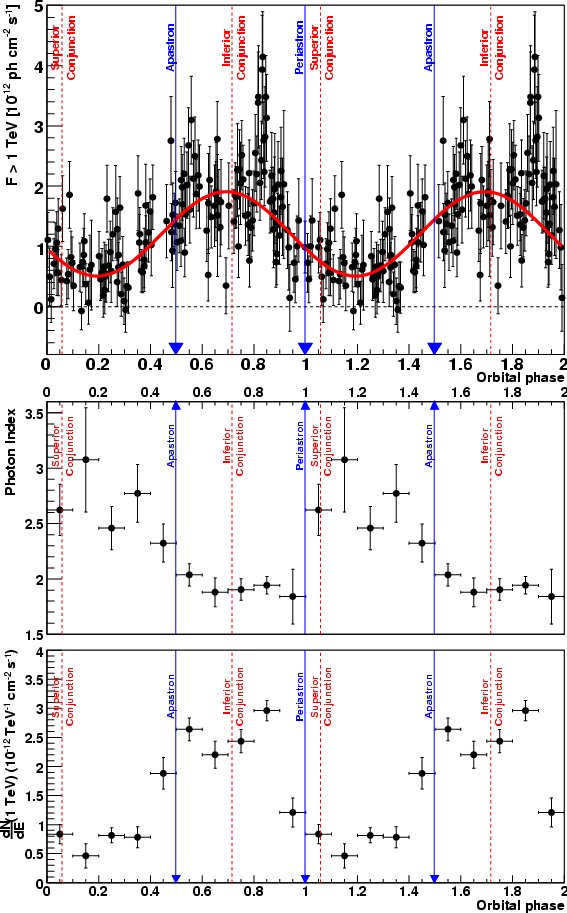}
\caption{The H.E.S.S. flux (bottom) and photon index (top) for LS~5039 as a function
of orbital phase. Figure from Aharonian et al. \cite{Aharonian_LS5039}.}
\label{HESS_LS5039_lc}
\end{figure}

Observationally, LS~5039 represents a rather challenging target for
Fermi-LAT, since it is located in a region with a large diffuse
background flux and with significant source confusion (notably from
PSR~J1826-1256). Results are first being shown at this conference
\cite{dubois_LS5039, abdo09_LS5039}, and reveal that the source is
detected at all orbital phases, with the emission peaking close to
\textit{superior} conjuction, in apparent anti-phase with the VHE
results. Spectral variability is also seen in the LAT data, with
$\Gamma=2.25\pm0.11$ at inferior conjuction and $\Gamma=1.91\pm0.16$
at superior conjuction. Most strikingly, a sharp spectral cut-off at
$E_o=1.9\pm0.5\U{GeV}$ is observed in the hard state, indicating that
the VHE spectra cannot be simply a smooth extrapolation of the lower
energy emission.

\begin{figure}
\centering
\includegraphics[width=75mm]{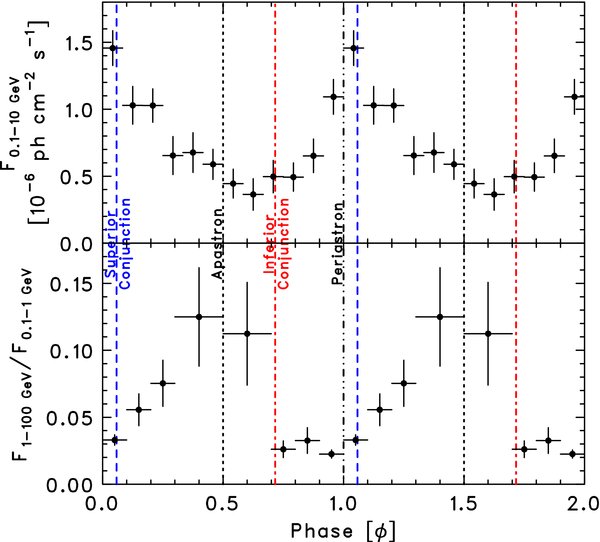}
\caption{The Fermi-LAT flux (top) and hardness ratio (bottom) for
LS~5039 as a function of orbital phase. Figure from Abdo et
al. \cite{abdo09_LS5039}.}
\label{Fermi_LS5039_lc}
\end{figure}

\subsection{LS~I~+61${^\circ}$303}

\begin{figure}
\centering
\includegraphics[width=75mm]{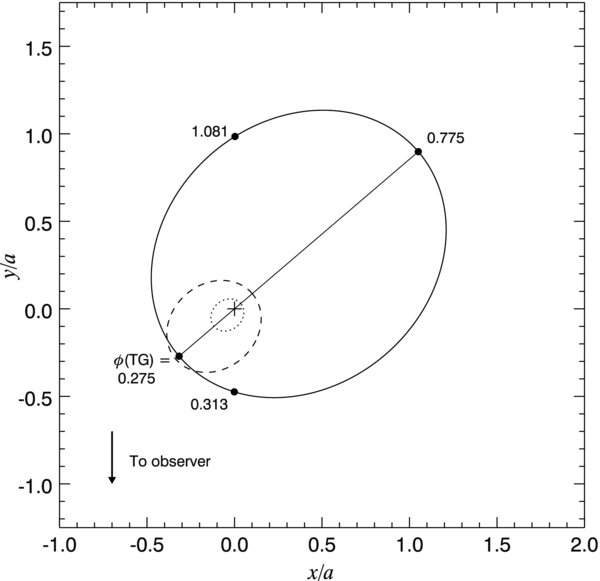}
\caption{A top view of the orbit of LS~I~+61${^\circ}$303, showing the
relative orbits (r/a) of the optical star and its compact companion of
unknown mass. The relative orbit of the compact object is shown as a
solid line, while the Be star's relative orbit depends greatly on the
mass of the companion. The dashed line indicates the Be star's orbit
assuming a 4 M$_{\odot}$ black hole, while the dotted line assumes a
1.4 M$_{\odot}$ neutron star. Figure from Aragona et
al. \cite{aragona09}.}
\label{LSIorbit}
\end{figure}

Similar to LS~5039, LS~I~+61${^\circ}$303 consists of a compact
object, either neutron star or black hole, in this case orbiting a
B0Ve star with a circumstellar disk ($\sim$12.5~M$_\odot$) in a 26.5
day orbit. The orbital eccentricity, $e$, is 0.537, and the orbital
separation varies from $\sim0.1\U{A.U.}$ at periastron ($\phi=0.275$)
to $\sim0.7\U{A.U.}$ at apastron. Figure~\ref{LSIorbit} sketches the
orbit of the system.

The original identification of a bright high energy source with COS-B
\cite{hermsen77}, coincident with a periodic radio source
\cite{gregory79}, ensured that LS~I~+61${^\circ}$303 would be a prime
target for EGRET. An EGRET source was duly detected \cite{tavani98},
showing some evidence for variability, but no measurable periodicity,
or correlation with emission at other wavelengths. This fact, along
with the large positional uncertainty, prevented a definitive
association of the HE source with the binary system. As with LS~5039,
evidence for radio jet structures has beeen found in
LS~I~+61${^\circ}$303 \cite{massi01}. More recent observations
question this microquasar interpretation \cite{dhawan06}, since the
radio structure can be seen to vary around the orbit (although see
\cite{romero07} for a counter-argument).

\begin{figure}
\centering
\includegraphics[width=75mm,trim=0 0 1mm 0, clip=true]{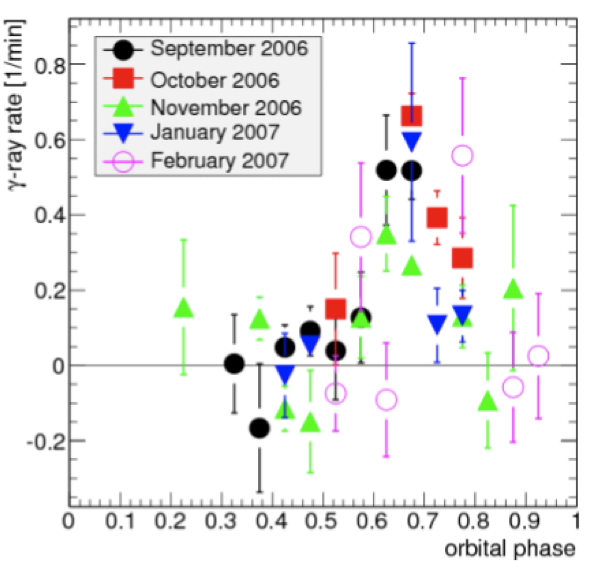}
\caption{The VERITAS gamma-ray rate from LS~I~+61${^\circ}$303 as a function
of orbital phase over a series of orbits.}
\label{LSI_VERITAS_lc}
\end{figure}

The detection of a variable VHE source at the location of
LS~I~+61${^\circ}$303 with MAGIC \cite{Albert_LSI_detect}, later
confirmed by VERITAS \cite{Acciari_LSI_detect}, completed the
identification this source as a gamma-ray binary. The object is now
one of the most heavily observed locations in the VHE sky, with deep
exposures by the two observatories spread over half a decade. Despite
this, the VHE source is much less well-characterized than LS~5039,
owing to its relatively weak VHE flux, and an inconvenient orbital
period which closely matches the lunar cycle, making observations over
all orbital phases almost impossible within a single observing
season. VHE emission is only clearly detected close to apastron (Figure~\ref{LSI_VERITAS_lc},
between phases $\phi=0.5-0.8$, with a peak flux $\sim10\%$ of the
steady Crab nebula flux and a power-law spectrum with index
$\Gamma\sim2.6$. Observations with MAGIC have established modulation
at the binary period \cite{Albert_LSI_period}, as well as a
correlation between VHE and XMM X-ray fluxes during 60\% of one orbit
\cite{anderhub09}. The only published VHE observations since the
launch of Fermi show no clear detection of the source, although
exposure close to the apastron phases was limited \cite{holder09}.

X-ray modulation at the orbital period has been measured with the
Rossi X-Ray Timing Explorer (RXTE) All-Sky Monitor \cite{leahy01,
wen06}, although regular observations taken every 2 days over a period
covering 6 orbital cycles in 2007-2008 with the RXTE Proportional
Counter Array (PCA) show no evidence for phase-dependent flux
modulation \cite{smith_LSI}. These same measurements, as well as later
measurements by Ray \& Hartman \cite{rayatel} reveal bright X-ray
flaring episodes lasting longer than 10 minutes, with substructure on
the timescale of a few seconds. A much shorter timescale flare,
lasting 0.23 seconds, has also been detected with the Swift Burst
Alert Telescope (BAT) \cite{dubusatel}. In both cases, the X-ray
variability may be related to other sources in the same field-of-view
(e.g. \cite{munozatel}). As with LS~5039, a hard X-ray spectrum and
absence of X-ray emission lines favours a non-thermal origin for the
emission.

\begin{figure}
\centering
\includegraphics[width=75mm]{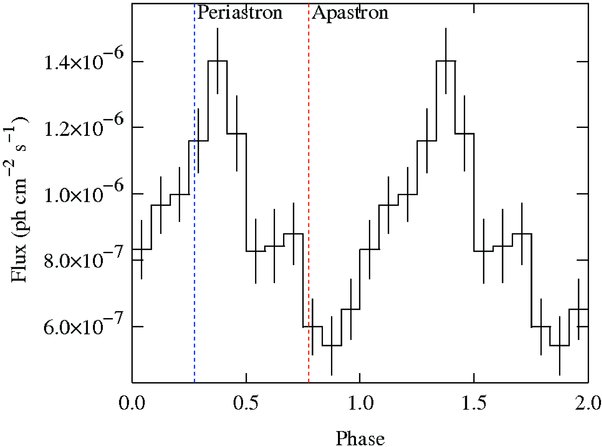}
\caption{The Fermi-LAT flux for LS~I~$+61^{\circ}$303 as a function of orbital phase. Figure from Abdo et
al. \cite{abdo09_LSI}.}
\label{LSI_Fermi_lc}
\end{figure}

\begin{figure}
\centering
\includegraphics[width=75mm]{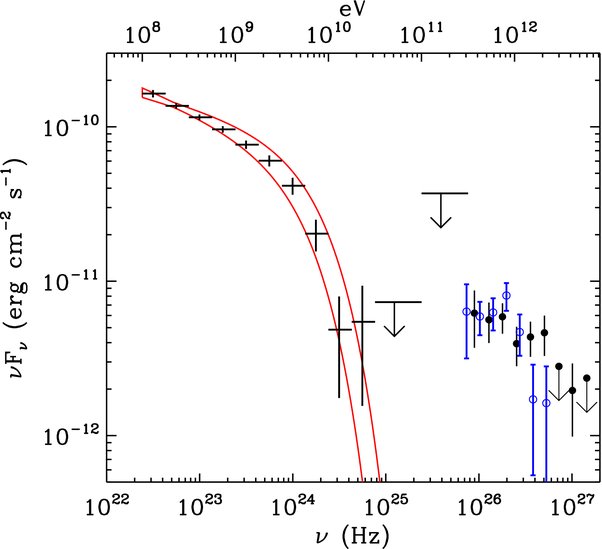}
\caption{Fitted spectrum of LS I +61°303 to the phase-averaged Fermi
data. The solid red lines are the ±1\u03c3 limits of the Fermi cutoff
power law; blue (open circle) data points from MAGIC (high state
phases 0.5-0.7); black (filled circle) data points from VERITAS (high
state phases 0.5-0.8). Data points in the Fermi range are likelihood
fits to photons in those energy bins. Note that the data from the
different telescopes are not contemporaneous, though they do cover
multiple orbital periods. Figure from Abdo et al. \cite{abdo09_LSI}.}
\label{LSI_Fermi_spec}
\end{figure}

LS~I~$+61^{\circ}$303 was one of the few non-pulsar objects firmly
identified in the Fermi-LAT Bright Source List, on the basis of the
orbital modulation of the gamma-ray flux \cite{abdo09_BSL}, and was
the only Galactic source on the first year LAT monitored source
list. Figure~\ref{LSI_Fermi_lc} shows the light-curve phase-averaged
over multiple orbits. The HE emission peaks close to periastron, in
apparent anti-phase with the non-contemporaneous VHE results. The LAT
spectrum is constant, within a power law index of
$\Gamma=2.21\pm0.04_{\mathrm{stat}}\pm0.06_{\mathrm{syst}}$ and, as
with LS~5039, displays sharp cut-off, at $E_o=6.3\pm1.1\U{GeV}$ (Figure~\ref{LSI_Fermi_spec}).

\subsection{Cygnus X-1}

As will be discussed in more detail later, PSR~B1259-63/SS~2883,
LS~5039 and LS~I~+61${^\circ}$303 can all plausibly be explained as
binary systems containing a non-accreting neutron star. The compact
object in the Cygnus X-1 system, however, is the best known candidate
for a stellar mass black hole. The system is believed to comprise a
$21\pm8$~M$_\odot$ black hole in a circular orbit around an O9.7~Iab
companion of $40\pm10$~M$_\odot$, with a period of 5.6
days\cite{gies86, ziolkowski05}. Cygnus X-1 is one of the brightest
known X-ray sources, and displays the well-known high/soft and
low/hard spectral states \cite{esin98}, which are believed to relate
to the accretion rate onto the compact object. The soft component is
observed when thermal emission from the accretion disk dominates,
while the high energy hard component is believed to be produced by
inverse Compton boosting, either by thermal electrons in a corona, or
at the base of a relativistic jet.

\begin{figure}
\centering
\includegraphics[width=75mm]{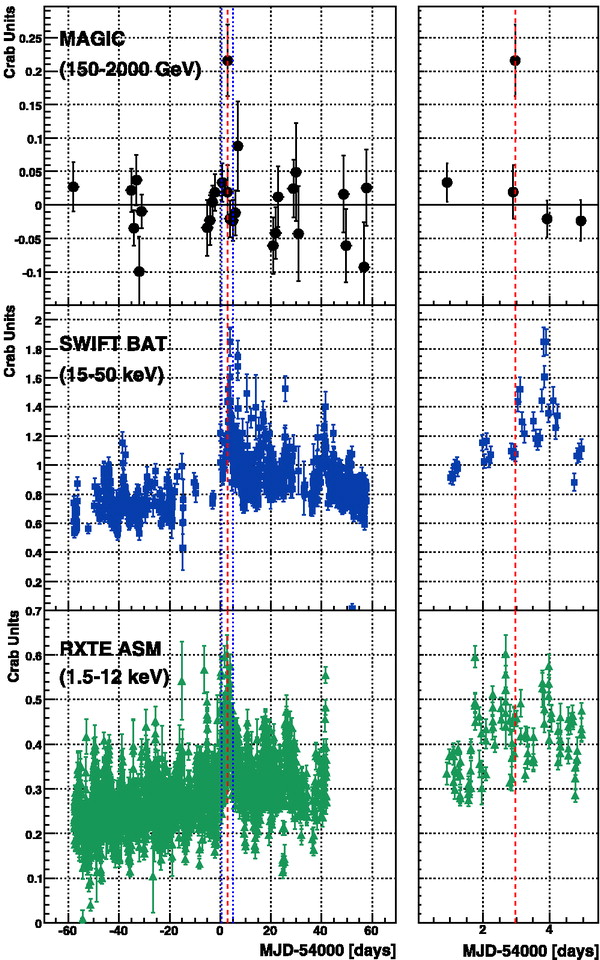}
\caption{MAGIC, Swift/BAT, and RXTE/ASM measured fluxes from Cygnus
X-1 as a function of time. The left panels show the whole time spanned
by MAGIC observations. The vertical, dotted blue lines delimit the
range zoomed in the right panels. The vertical, dashed red line marks
the time of the MAGIC signal. Figure from Albert et
al. \cite{albert07_CygX1}.}
\label{CygX1_MAGIC_lc}
\end{figure}

Gamma-ray emission from Cygnus~X-1 has not been detected by any
satellite observatories. In the VHE regime, a deep, $40\U{hour}$
exposure by MAGIC in 2006 also showed no evidence for steady emission;
however, a search for variability on shorter timescales revealed an
excess signal with a statistical significance of $4.1\sigma$ during a
single 79 minute period \cite{albert07_CygX1}. As illustrated in
Figure~\ref{CygX1_MAGIC_lc}, the VHE excess occurred during the rising
edge of a hard X-ray flare; observations on the following night during
the falling edge of a second X-ray peak showed no such excess. The
result is intriguing, being the first evidence for VHE emission from
an accreting binary system, but clearly requires follow-up
observations for confirmation and characterization of the gamma-ray
light curve structure before any strong conclusions can be drawn.

\subsection{Cygnus X-3}

\begin{figure*}
\centering
\includegraphics[width=110mm]{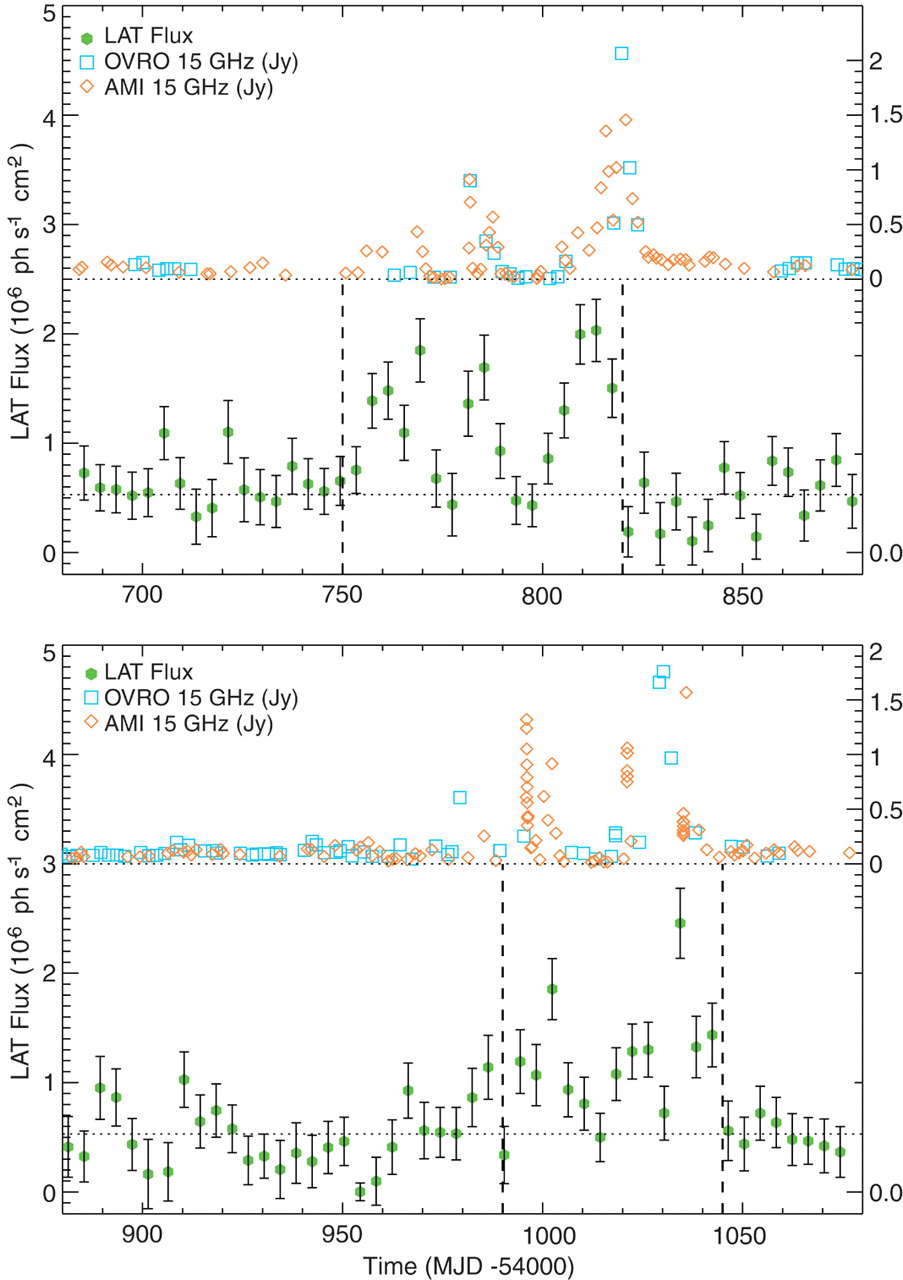}
\caption{Fermi-LAT gamma-ray light curve in two parts (top and
 bottom).  The vertical dashed lines illustrate the active periods
 corresponding to a soft X-ray state. The Cyg X-3 15-GHz radio flux
 (from the AMI and OVRO 40-m radio telescopes) is shown at the top of
 each panel. Figure from Abdo et al. \cite{fermi09_CygX3}.}
\label{Fermi_CygX3_lc}
\end{figure*}

Cygnus X-3 is another presumably accretion-powered system, comprised
of a 10-20~M$_\odot$ compact object and a Wolf-rayet star companion,
with a short orbital period of $4.8\U{hours}$ \cite{parsignault72}.
As discussed in the introduction, Cygnus~X-3 holds an important place
in the history of the field of gamma-ray astronomy, particularly for
the ground-based instruments. At present, there is no strong evidence
for VHE emission from the source. Compelling evidence for HE emission
was also lacking until very recently, with the appearance of new
results from AGILE \cite{tavani09_CygX3} and Fermi-LAT
\cite{fermi09_CygX3}. The results are discussed in greater detail
elsewhere in these proceedings \cite{corbel09}. Briefly, both the LAT
and AGILE observe HE emission only intermittently, during soft X-ray
states in which radio flares are also observed. These radio flares are
believed to be associated with relativistic plasma ejection events,
and the peak gamma-ray emission appears to precede major radio flares
by a few days. Figure~\ref{Fermi_CygX3_lc} shows the LAT lightcurve
for Cygnus X-3. The emission seen by the LAT during the active phases
is also observed to be modulated at the $4.8\U{hour}$ orbital period,
providing the definitive association of the gamma-ray source with
Cygnus~X-3.

\subsection{An Enigma: HESS~J0632+057}

The sources discussed above are all well-known objects at radio
through X-ray wavelengths and their binary nature is certain. The
reported gamma-ray observations clearly show that binary systems can
be among the brightest objects in the GeV-TeV sky, and so the
possibility of identifying new binaries through their high energy emission
alone is not unreasonable. Hinton et al. \cite{hinton09} have proposed
that one of the unidentified TeV sources discovered by H.E.S.S.,
HESS\,J0632+057 \cite{aharonian07_monoceres}, might be an example of
such a gamma-ray binary system. This source was originally detected during two
observations of the Monoceros Loop SNR region, separated by
$1\U{year}$, with integral fluxes above $1\U{TeV}$ of $\sim3\%$ of the
Crab Nebula flux. At TeV energies, binary systems are
indistinguishable from point-like objects; of the $\sim40$
unidentified TeV sources, only HESS\,J0632+057 and the Galactic centre
source HESS\,J1745-290 are point-like. Follow-up observations with
XMM-Newton have revealed an X-ray source, XMMU\,J063259.3+054801 coincident
with the TeV source, and with a massive star MWC\,148, spectral type
B0pe. The X-ray source presents a hard power-law spectrum
($\Gamma=1.26\pm0.04$) and is variable on hour timescales - consistent
with the X-ray properties of established gamma-ray binary systems.

\begin{figure*}
\includegraphics[width=150mm, trim=0 12mm 0 0, clip=true]{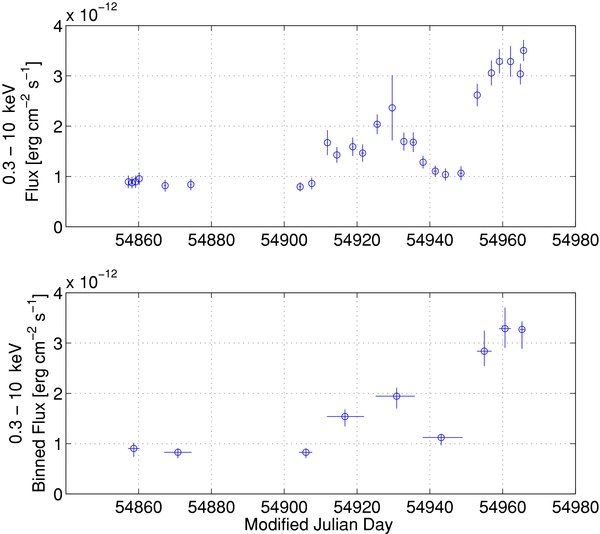}
\caption{ Swift-XRT observations of XMMU~J063259.3+054801 in the
 0.3-10 keV band.  Figure from Falcone et al. \cite{falcone10}.}
\label{Swift_0632_lc}
\end{figure*}

\begin{figure}
\centering
\includegraphics[width=75mm]{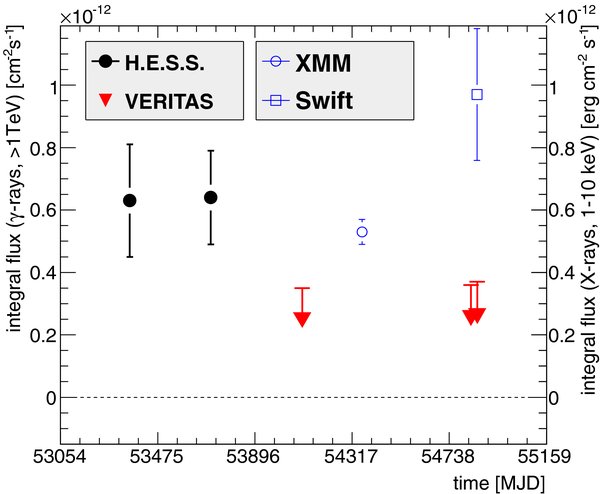}
\caption{Light curve above $1\U{TeV}$ for HESS~J0632+057. The downward
 pointing arrows show the 99\% confidence limits derived from the
 VERITAS data. The X-ray fluxes measured by XMM-Newton and Swift are
 indicated by open symbols.  Figure from Acciari et
 al. \cite{acciari_0632}.}
\label{VERITAS_0632_lc}
\end{figure}

The known gamma-ray binary systems have already undergone many years
of intensive multiwavelength study, which has allowed to construct
full spectral energy distributions, and to characterize the periodic
and non-periodic variability timescales. Such studies are now just
beginning for HESS\,J0632+057. A faint, unresolved radio source has
been detected at the position of MWC\,148. The VLA observations
\cite{skilton09} consist of three exposures separated by $\sim1$ month
and show significant variability at $5\U{GHz}$. Ongoing observations
with XRT onboard the Swift satellite reveal X-ray variability on
timescales from days to months, with no clear evidence for periodicity
as yet (Figure~\ref{Swift_0632_lc}). Follow-up VHE observations by
VERITAS resulted in upper limits below the reported H.E.S.S. flux
(Figure~\ref{VERITAS_0632_lc}), indicating that the source is also
variable at TeV energies \cite{acciari_0632}. At present, there is no
known HE counterpart to HESS\,J0632+057; the closest source is
0FGL\,J0633.5+0634 (LAT PSR J0633+06), at an angular separation of
$0.8^\circ$.

As things stand, the nature of HESS\,J0632+057 remains a mystery. As a
variable, point-like TeV source, co-located with a massive Be star, it
is certainly a promising binary candidate, and the overall spectral
energy distribution is compatible with this scenario. Contemporaneous
measurements over all wavelengths will help to shed light on the
situation, and the detection of a periodic emission component at any
wavelength would provide definitive evidence.

\section{Discussion}

Interpretation of the emission from Galactic gamma-ray binary systems
drives an extremely active field, a full review of which is beyond the
scope of this article. An overview of some of the important issues,
and a selection of the different models can be found in these articles
and references therein:\cite{romero05, bosch06, dubus06, gupta06,
orellana07, sierpowska09, chernyakova09}.

One of the key questions to answer is the ultimate nature of the power
source: whether accretion onto a compact object, or the spin-down
power of a neutron star. In the case where pulsar emission is
observed, such as with PSR~B1259-63/SS~2883, the answer is clear. For
the other sources, the question remains, although the evidence at
present (in the opinion of this reviewer) favours LS~5039 and
LS~I~+61${^\circ}$303 as non-accreting, pulsar-wind driven systems
(e.g. \cite{dubus06}), and Cygnus X-3 and Cygnus X-1 as accreting
systems driving relativistic jets. Too little is known about
HESS~J0632+057 to make even an educated guess.

Beyond this basic paradigm, the details of the interpretation can vary
widely. The particle acceleration mechanism may be through colliding
shocks in a jet, magnetic reconnection events, or shock acceleration
at the pulsar wind - stellar wind interaction. The energetic particle
population may be hadron or lepton dominated, leading to different
production mechanisms for the high energy emission: $\pi^0$-decay
through hadronic interactions with surrounding matter, or inverse
Compton boosting of local photon fields by energetic leptons. The
gamma-ray emission can originate in very different locations: close to
a jet or in the circumstellar environment, in the wind interaction
region or the pulsar wind zone or, in the case of curvature radiation,
within the pulsar magnetosphere itself. Once produced, the observed
gamma-ray flux is modulated by a number of factors: both the geometry
of the system with respect to the line of sight, and varying photon
field intensities and matter density around the orbit lead to varying
energy-dependent gamma-ray production and absorption
efficiencies. Additionally, a number of other effects, such as stellar
wind clumping, the impact of geometrical uncertainties and particle
cascading could play a significant role.

\begin{figure*}
\centering
\includegraphics[width=165mm]{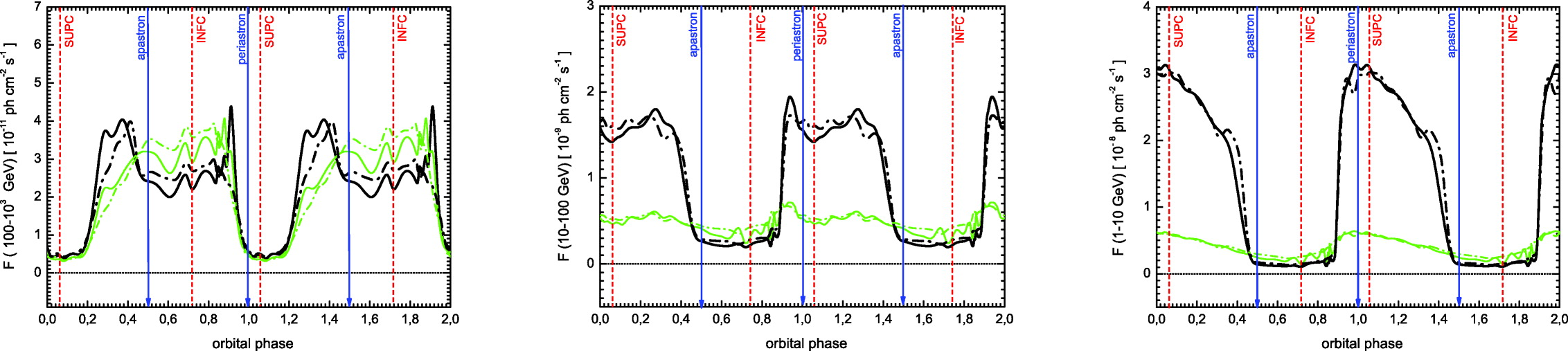}
\caption{Predicted theoretical light curves for the energy intervals
100 GeV-1 TeV, 10-100 GeV, and 1-10 GeV. Both inclination angles and
interacting lepton spectra considered are shown (dot-dashed curves, i
= 30°; solid curves, i = 60°; black curves, variable spectrum of
primary leptons; green curves, constant spectrum along the
orbit).Figure from \cite{sierpowska08}}
\label{model}
\end{figure*}

Despite this wide range of factors to address, significant progress
has been made in the efforts to model these systems, and concrete
predictions exist which are now being put to the test by the Fermi-LAT
observations and the TeV instruments. To give just one example,
Figure~\ref{model} shows predicted light curves over the HE and VHE
energy bands for LS~5039, in the context of a leptonic particle
population powered by a non-accreting pulsar wind \cite{sierpowska08}. 

In the near future, we can expect a number of observational questions
to be addressed. The existence of the Fermi-LAT GeV cut-offs in
LS~5039 and LS~I~+61${^\circ}$303 is particularly intriguing, since it
implies that the HE and VHE emission may have different origins (for
example, the HE emission could be magnetospheric, or associated with a
hadronic particle population). The identification of a pulsed
component to the gamma-ray emission would clarify this. It seems
likely that the LAT will also add to the catalog of gamma-ray
binaries, and this information can be used to guide new VHE
observations: Cygnus X-3 is already highlighted as an obvious
candidate for targeted VHE follow-up. The strongest observational
constraints to the emission models are always provided by long-term,
strictly contemporaneous multi-wavelength campaigns, several of which
are presently underway or planned for the near future. The next
periastron pass for PSR~B1259-63/SS~2883 at the end of 2010 will be
the first chance to observe both the HE and VHE emission from this
object while, at the time of writing, the Be-pulsar binary 1A~0535+262
has entered a giant outburst state. This occurs only once every five -
ten years, and so represents the first opportunity for sensitive
HE/VHE observations of this object.

\bigskip 
\begin{acknowledgments}

The author wishes to thank the organisers for the invitation to speak
at the Symposium, and Richard Dubois and Matthias Kerschhaggl for
their help during the preparation of the accompanying presentation.

Work supported, in part, by the NASA Fermi Cycle 2 Guest Investigator
Program.

\end{acknowledgments}

\newpage
\bigskip 

\end{document}